\PassOptionsToPackage{square, numbers, sort}{natbib}
\documentclass[amssymb,aps,preprint,pra]{revtex4-2}
\usepackage[utf8]{inputenc}
\usepackage[T1]{fontenc}
\usepackage[english]{babel}

\usepackage[fleqn]{amsmath}
\usepackage{mathtools}
\usepackage{amssymb}
\usepackage{xfrac}
\usepackage{placeins}
\usepackage{graphicx}
\usepackage[percent]{overpic}
\usepackage{lineno, blindtext}

\newcommand{\ii}{\mathrm{i}}  
\newcommand{\ee}{\mathrm{e}}  

\usepackage{xcolor}
\usepackage[normalem]{ulem}

\begin{document}

\title{Wigner time delay and Hartman effect in quantum motion along deformed Riemannian manifolds}
\author{Benjamin Schwager, Lars Meschede, Jamal Berakdar}
\affiliation{Institut für Physik, Martin-Luther-Universität Halle-Wittenberg}
\date{\today}
\begin{abstract}
 Elastic scattering of a wave can be quantified by a shift in the phase with respect to the incoming wave phase. A qualitative measure of the time during which the effect occurs is given by the Wigner time delay. The tunneling time in turn is known to saturate with increasing tunneling barrier width (Hartman effect). Here, we analyze the elastic quantum mechanical scattering in a deformed one-dimensional Riemannian manifold, particularly with respect to the Wigner time delay and conclude on the Hartman effect. It is shown that scattering due to local curvature variations imply imperfect conduction behavior indicating resonance states and leads to a Wigner time delay which, at low energies, is in variance with the classical time delay that is inferred from the arc length. At moderate and high energies, however, classical and quantum time delays coincide.
\end{abstract}

\maketitle
\section{Introduction}
For coherent elastic wave scattering, the phase of the wave remains fully matched during its spatial propagation. When the wave traverses a localized scatterer, its phase gradually shifts and remains unchanged in absence of further scattering events. The so-attained wave-vector-dependent phase shift encodes complete information on the scattering process \citep{Joachain1987}. In particular, from the energy-dependence of the phase shift that a monochromatic wave experiences one may extract a quantity which qualitatively relates to the duration of scattering, i.e., the scattering time or the (Wigner) time delay of the signal upon scattering. Following the original derivation of Wigner \citep{Wigner1955} for elastic scattering of a narrow-banded wave packet, one may view the Wigner time delay from a wave propagation perspective, demanding stationariness of the phase relations between the components in both the incoming and the scattered waves.
In particular, considering an incident wave with a well-defined wave vector $\textbf{k}$ and energy $E$, its initial phase is $\Phi_{\mathrm{i}} = kr - E(k) t/\hbar$. Stationariness of the total phase $\Phi(k) = \Phi_{\mathrm{i}}(k) + \Delta \Phi(k)$, where $\Delta\Phi(k)$ is the phase shift caused by scattering, incorporates the derivative $\partial_k\Phi(k)=\partial_k(kr -E t/\hbar + \Delta\Phi(k)) =  r - \frac{\partial_{k}(E)}{\hbar}\,\left( t - \hbar\,\frac{\partial_{k}\Delta\Phi}{\partial_{k}E} \right)$.
Therein, we interpret the quantity
$
    \tau = \hbar\,\frac{\partial}{\partial E}\Delta \Phi
$
as the time (shift) or delay brought about  by the scattering process.
\par
The relation between this time and the tunneling time has been the subject of a long-lasting discussion, which is commented in detail, for example, in \citep{Hauge1989, Rivlin2021}. For an electron receding from a parent target upon photoionization, the Wigner time delay has been measured and analyzed theoretically \citep{Schultze2010, Sainadh2019, Banerjee2020, Kheifets2013, Waetzel2014,PhysRevA.94.033414,Yu2022}. Our focus here lies in the analysis of scattering processes triggered by variations in the geometry of space. Recently, we discussed the motion of a quantum particle in non-flat spaces \citep{Meschede2023}. We found that localized curvature modulations in Riemannian manifolds act as a source for quantum scattering. It has been shown that perturbation methods exhibit anomalous behavior compared to established knowledge in flat spaces. Transport through deformed quantum wires was considered, for example, in \citep{Taira2007, Zhang2007, Serafim2021}, incorporating electromagnetic fields and spin-orbit coupling. Investigations on how the time delay behaves in this case have not been studied yet, and another open question is whether a Hartman effect \citep{Hartman1962} exists, i.e., whether the delay time saturates and becomes constant when the extension of the scattering barrier (for a fixed height) is increased.
\par
To clarify these issues we formulated and simulated the quantum scattering of a quantum particle freely moving on a one-dimensional Riemannian manifold. In the asymptotic region, the manifold converges to a straight line (flat space).No further assumptions are put on the internal structure of the space, so electron-phonon coupling and the role of polaronic effects are not considered in the model. The S-matrix is evaluated with spectral resolution and both the Wigner time delay and the transmission coefficient are derived. It is confirmed that the nontrivial geometric properties of the underlying space cause quantum wave scattering that can be quantified by the behavior of the transmission coefficient. Besides, a corresponding classical particle does not experience any scattering. The time delays for both physical regimes are contrasted. To study possible interference effects, we consider two Gaussian dents with a varying distance and demonstrate the sensitivity of the results with respect to these shapes. Our findings support that deformations have a significant influence on transport experiments.
\par
The article is organized as follows: In Sec. \ref{sec:Background} we formulate our model for a nonrelativistic spinless quantum particle within a one-dimensional Riemannian manifold based on the confinement potential approach \citep{Jensen1971, Costa1981,Meschede2023}. We present and discuss the results of our scattering simulations in Sec. \ref{sec:Results}. Finally, Sec. \ref{sec:Conclusion} summarizes the findings.

%
%
%
\section{Formulation of Quantum Mechanics on the Wire}
\subsection{Analytical model}
\label{sec:Background}
We study a one-dimensional Riemannian manifold $(\mathcal{M}, g)$ that is isometrically embedded in the ambient three-dimensional space (hyperspace). The canonically induced Riemannian metric tensor field and the Levi-Civita connection being imprinted on the tangent bundle $T\mathcal{M}$ are assumed. The manifold is parametrized in Monge form as
\begin{align}
    \mathcal{Y}: \mathbb{R}^{1}\supset \mathcal{Q} \rightarrow \mathcal{M}\subset \mathbb{R}^{3}\ ,\ q^{1} \mapsto \mathcal{Y}(q^{1}) \coloneqq \left( \begin{matrix}
        q^{1} \\
        f(q^{1}) \\
        0
    \end{matrix} \right)\ ,
\label{eq:curve_parametrization}
\end{align}
where we assume $f\in\mathcal{C}^{\infty}(\mathbb{R})$ (although this could be loosened) and that it decays sufficiently fast with respect to a center $x_{0}\in\mathbb{R}$. It thus represents an asymptotically flat planar curve. A spinless charged quantum particle with mass $m_{\rm 0}$ is confined to this space and hence becomes subject to geometry-induced effects. We apply the confinement potential approach (CPA) \citep{Jensen1971, Costa1981} to confine the motion to $(\mathcal{M}, g)$ in the presence of the hyperspace.
\par
The essence of this procedure is that \citep{Costa1981, Maraner1995, Schuster2003, Jalalzadeh2005, Brandt2017} the particle is constrained to the manifold by some (in our case scalar) confinement potential $V_{\lambda}$ with the following properties: (i) $V_{\lambda}$ depends only on the normal displacement coordinate, (ii) $V_{\lambda}$ has a deep minimum on $\mathcal{M}$, and (iii) $V_{\lambda}$ preserves the gauge group of the resulting dimensionally reduced theory, which has to be a subgroup of the isometry group of the ambient space. Then, the parameters that describe position in space are divided into the extrinsic and intrinsic ones, describing the tangent and normal degrees of freedom, respectively, and ultimately the Schrödinger equation can be decoupled into two which are valid for the effective dynamics. Note that with condition (ii) the uncertainty principle of Heisenberg is maintained, and that (iii) has a major influence on the resulting dynamics as the confinement may give rise to geometry-induced Yang-Mills fields and particle properties \citep{Maraner1993, Maraner2008, Brandt2017, Gravesen2018, Wang2018}. Here, we restrict ourselves to tubular confinements so that the effective dynamics possesses a $\mathrm{SO}(2)$ gauge symmetry.
\par
Assuming the arc length parametrization, the resulting effective external Schrödinger equation reads \citep{Costa1981}
\begin{align}
    \ii\hbar\,\partial_{t}\chi_{\mathrm{t}}(s) = \hat{T}[g,\kappa]\chi_{\mathrm{t}}(s) = -\frac{\hbar^{2}}{2m_{0}}\,\left[ \partial_{s}^{2} + \frac{\kappa^{2}(s)}{4}\,\hat{I} \right]\,\chi_{\mathrm{t}}(s)\ ,
\label{eq:Schrodinger_arc}
\end{align}
where the term
\begin{align}
    V_{\mathrm{geo}}(s) = -\frac{\hbar^{2}}{8m_{0}}\,\kappa^{2}(s)
\label{eq:geometric_Pot}
\end{align}
is commonly called the geometric potential. Actually, it is better understood as a correction term to the operator of kinetic energy to account for the geometric properties of the Riemannian manifold \citep{Liu2007, Liu2011, Wang2017}. In this case it holds that $\kappa = \Vert \partial_{s}\textbf{t} \Vert_{2}$, but in practice this concrete parametrization might not be obvious. Here, we work with the one set in \eqref{eq:curve_parametrization}. Re-parametrizing the above equation and inserting the general formula of curvature \citep{Carmo2016},
\begin{align}
    \kappa = \frac{\Vert \textbf{t} \times \textbf{t}_{,1}\Vert_{2}}{\Vert \textbf{t} \Vert_{2}^{3}}\ ,\ \textbf{t} = \partial_{1}\mathcal{Y} = \mathcal{Y}_{,1}\ ,
\end{align}
we find
\begin{align}
\begin{aligned}
     \ii\hbar\,\partial_{t}\chi_{\mathrm{t}}(q^{1}) &= \hat{T}[g,\kappa]\,\chi_{\mathrm{t}}(q^{1}) \\
     &= -\frac{\hbar^{2}}{2m_{0}}\,\left[ \frac{1}{1+ f_{,1}^{2}}\partial_{1}^{2} - \frac{f_{,1}\,f_{,11}}{\left( 1+ f_{,1}^{2} \right)^{2}}\partial_{1} + \frac{1}{4}\,\frac{f_{,11}^{2}}{\left(1+ f_{,1}^{2}\right)^{3}}\,\hat{I} \right]\,\chi_{\mathrm{t}}(q^{1}) \\
     &=  -\frac{\hbar^{2}}{2m_{\mathrm{eff}}}\,\left[ \partial_{1}^{2} - \frac{f_{,1}\,f_{,11}}{1+ f_{,1}^{2}}\partial_{1} + \frac{1}{4}\,\frac{f_{,11}^{2}}{\left(1+ f_{,1}^{2}\right)^{2}}\,\hat{I} \right]\,\chi_{\mathrm{t}}(q^{1})\ .
\end{aligned}
\label{eq:Schrodinger_graph}
\end{align}
Here, we introduced the spatially varying effective mass
\begin{align}
    m_{\mathrm{eff}} \coloneqq (1+ f_{,1}^{2})\,m_{0}\ .
\label{eq:effective_mass}
\end{align}
Both the wave functions solving \eqref{eq:Schrodinger_arc} and \eqref{eq:Schrodinger_graph} are equivalent and describe a nonrelativistic spinless quantum particle travelling freely along the curved manifold. Therein, the operator of kinetic energy is expressed as a functional of the geometric invariants of the configuration space $(\mathcal{M}, g)$, which are the metric tensor field $g$ and the curvature field $\kappa$. In this way, the quantum dynamics are effected by the spatial geometry, leading to scattering. Mathematically, the CPA results in  a more general Sturm-Liouville problem than the Schrödinger operator in Euclidean space (hyperspace). As such, the existence of an orthogonal (with respect to the parameter space) system of eigenfunctions with corresponding unique real eigenvalues is guaranteed \citep{Zettl2005}, so that the scattering-experiment is well-defined.
\par
We can express the problem differently and simplify the equations by making the ansatz $\chi_{\mathrm{t}}(q^{1}) \doteq \alpha(q^{1})\,\psi_{\mathrm{t}}(q^{1})$, which leads to
\begin{align}
    \ii\hbar\,\partial_{t}\psi_{\mathrm{t}} = -\frac{\hbar^{2}}{2m_{\mathrm{eff}}}\,\frac{1}{\alpha}\hat{L}[\alpha]\,\psi_{\mathrm{t}}
\label{eq:sep_Schrodinger}
\end{align}
with the operator
\begin{align}
    \frac{1}{\alpha}\hat{L}[\alpha] \coloneqq \partial_{1}^{2} + \left( 2\,\frac{\alpha_{,1}}{\alpha} - \frac{f_{,1}\,f_{,11}}{1 + f_{,1}^{2}} \right)\,\partial_{1} + \left( \frac{\alpha_{,11}}{\alpha} - \frac{f_{,1}\,f_{,11}}{1 + f_{,1}^{2}}\,\frac{\alpha_{,1}}{\alpha} + \frac{1}{4}\,\frac{f_{,11}^{2}}{\left( 1+ f_{,1}^{2} \right)^{2}}  \right)\,\hat{I} \ .
\label{eq:L_operator}
\end{align}
Now, we can determine the function $\alpha$ such that the contribution of the first order derivative vanishes. 
To do so we write 
\begin{align}
     2\,\frac{\alpha_{,1}}{\alpha} - \frac{f_{,1}\,f_{,11}}{1 + f_{,1}^{2}} \overset{!}{=} 0 \ \Leftrightarrow\ \partial_{1}\,\mathrm{ln}(\alpha) = \frac{1}{4}\,\partial_{1}\,\mathrm{ln}\left( 1 + f_{,1}^{2} \right)\ ,
\end{align}
implying
\begin{align}
    \alpha = \sqrt[4]{1+f_{,1}^{2}} + \alpha_{0}\ ,\ \alpha_{0}\in\mathbb{R}\ .
\label{eq:alpha_function}
\end{align}
Combining \eqref{eq:sep_Schrodinger}, \eqref{eq:L_operator} and \eqref{eq:alpha_function}, we can state the effective external Schrödinger equation as
\begin{align}
    \ii\hbar\,\partial_{t}\psi_{\mathrm{t}} = \left[ -\frac{\hbar^{2}}{2m_{\mathrm{eff}}}\,\partial_{1}^{2} + V_{\mathrm{eff}}\,\hat{I} \right]\,\psi_{\mathrm{t}}\ .
\label{eq:effective_Schrodinger_effective}
\end{align}
In contrast to \eqref{eq:Schrodinger_graph},  Eq. \eqref{eq:effective_Schrodinger_effective} describes a particle moving along the $x$-axis, thereby possessing the position dependent mass \eqref{eq:effective_mass} and encountering the potential landscape
\begin{align}
\begin{aligned}
    V_{\mathrm{eff}} &\coloneqq -\frac{\hbar^{2}}{2m_{\mathrm{eff}}}\,\frac{-3\,\left( -1+f_{,1}^{2} \right)\,f_{,11}^{2} + 2\,\left( 1+f_{,1}^{2} \right)\,f_{,1}\,f_{,111}}{4\,\left( 1+f_{,1}^{2} \right)^{2}} \\
    &= \left[ 2\,\left( 1+f_{,1}^{2} \right)\,\frac{f_{,1}\,f_{,111}}{f_{,11}^{2}} + 3\,\left( 1 - f_{,1}^{2} \right) \right]\,V_{\mathrm{geo}}\ .
\end{aligned}
\label{eq:effective_Pot}
\end{align}
The above transformation re-states the problem. Instead of considering a motion along a curve through the two-dimensional plane, we can equivalently study a motion along a straight line (the $x$-axis). The drift along the second dimension(the $y$-axis) is incorporated in the above defined effective particle mass and potential field. It is interesting to note that the geometric potential assumes a traditional role within this effective Schrödinger equation \eqref{eq:effective_Schrodinger_effective}: While $V_{\mathrm{geo}}$ is generally not a quantum potential in the usual sense (rather it appears as a geometric correction term that complements  the operator of kinetic energy),  $V_{\mathrm{eff}}$ indeed acts as a scalar potential term in its own right as it is added to an already valid kinetic energy operator of a freely moving particle with varying mass.
\par
If we choose, for convenience, the boundary conditions such that $\alpha_{0} = 0$, we can infer the limiting behavior $\underset{\vert x\vert \rightarrow \infty}{\mathrm{lim}}\alpha(x) = 1$. Hence, both the equations \eqref{eq:Schrodinger_graph} and \eqref{eq:effective_Schrodinger_effective} and their respective solutions posses the same asymptotic form, whose solutions are plane waves. The general asymptotic solution is hence given as
\begin{align}
    \psi_{\mathrm{asy}} = C_{\mathrm{l}}\,\ee^{\ii k q^{1}} + C_{\mathrm{r}}\,\ee^{-\ii k q^{1}}\ ,\ C_{\mathrm{l}},C_{\mathrm{r}}\in\mathbb{C}\ ,
\end{align}
where the labels of the coefficients denote whether the respective part of the wave function travels away from the left ($\mathrm{l}$) or the right ($\mathrm{r}$) channel. Then, we can use equation \eqref{eq:effective_Schrodinger_effective} and proceed within the S-matrix formalism to investigate the geometric influences of the underlying space on the particle motion. We expect that due to scattering processes that occur upon localized geometric structures acting as short range targets, quantum transport through the wire shall be affected even if only free motion is incorporated. Let $A$ and $B$ denote the amplitude vectors of the incoming and outgoing states, respectively, then the S-matrix can be written as \citep{Texier2016}
\begin{align}
    B = \mathcal{S} A\ \Leftrightarrow\ \begin{pmatrix}
        B_{\mathrm{l}} \\
        B_{\mathrm{r}}
    \end{pmatrix} = \begin{pmatrix}
        r_{\mathrm{l}} & t_{\mathrm{r}} \\
        t_{\mathrm{l}} & r_{\mathrm{r}}
    \end{pmatrix} \cdot \begin{pmatrix}
        A_{\mathrm{l}} \\
        A_{\mathrm{r}}
    \end{pmatrix}\ .
\end{align}
From this, the reflection and transmission coefficients can be derived. For convenience, we restrict ourselves to incoming states from the left side ($A_{\mathrm{l}} = 1$ and $A_{\mathrm{r}} = 0$) and define the reflection coefficient $R$ and transmission coefficient $T$ as
\begin{align}
    R \coloneqq \vert \mathcal{S}_{11}\vert^{2}\ ,\ T \coloneqq \vert \mathcal{S}_{21}\vert^{2}\ ,
\label{eq:refelctivity+transmissivity}
\end{align}
fulfilling $T + R = 1$. These quantities describes the probability of measuring the particle entering the scattering region from the left channel in either the left or the right one, when it leaves.
\par
Furthermore, the data from the S-matrix can be used to derive scattering phase shifts and, consequently, statistical time delay parameters such as the Wigner time delay
\begin{align}
    \tau_{\mathrm{W}} = -\frac{\ii \hbar}{N}\,\mathrm{Tr}\left\{ \mathcal{S}^{\dagger}\frac{\partial}{\partial E}\mathcal{S} \right\} = \frac{\hbar}{N}\,\frac{\partial}{\partial E} \Phi_{\mathrm{F}}\ ,
\label{eq:Wigner_TD}
\end{align}
where
\begin{align}
    \Phi_{\mathrm{F}} = \mathrm{det}(\mathcal{S})
\end{align}
is the Friedel phase and $N = 2$ denotes the number of scattering channels. Mathematically, the Wigner time delay stems from a phase shift of an asymptotic state due to the presence of an interaction. With respect to a particle picture it may be interpreted as the duration that the projectile spends in the scattering region, that is, before being measured in either the reflection or the transmission channel. For comparison, we consider the duration that a classical particle, whose kinetic energy is given by the average incoming particle energy $E$, needs to pass through the wire of arc length $D\in\mathbb{R}^{+}$:
\begin{align}
    \tau_{\mathrm{c}} = \sqrt{\frac{m_{0}}{2E}}\,D\ .
\label{eq:classical_TD}
\end{align}
Clearly, a classical particle does not undergo any scattering/reflection event when moving freely in a manifold and can thus only be measured in the transmission channel.

\subsection{Implementation within the tight-binding model}
We study the unbound states of this landscape by means of stationary states. Therefore, we obtain numerical results by considering \eqref{eq:effective_Schrodinger_effective}, using the Kwant code \citep{Groth2014}. To this end, the Hamilton operator is expanded in the tight-binding basis $\vert q^{1}_{i}\rangle \doteq \vert i\rangle$, and only next-neighbor interaction is incorporated. Based on finite differences with lattice constant $a\in\mathbb{R}^{+}$, the representation in density operators then reads
\begin{align}
\begin{aligned}
    \hat{H} &= \sum_{i} \left[ H_{ii}\,\vert i\rangle\langle i\vert + H_{i+1\,i}\,\vert i+1\rangle\langle i\vert + H_{i\,i+1}\,\vert i\rangle\langle i+1\vert \right] \\
    &= \sum_{i}\left[ -t_{\mathrm{eff}}\left(q^{1}_{i} + \frac{a}{2}\right)\,\left(  \vert i+1\rangle\langle i\vert + \vert i\rangle\langle i+1\vert \right) + \left( V_{\mathrm{eff}}(q^{1}_{i}) +  2\langle t_{\mathrm{eff}}\rangle_{i} \right)\,\vert i\rangle\langle i\vert \right].
\end{aligned}
\end{align}
The hopping parameter of the particle with position-dependent mass is defined as
\begin{align}
    t_{\mathrm{eff}}(q^{1}) \coloneqq \frac{\hbar^{2}}{2 m_{\mathrm{eff}}(q^{1})\,a^{2}} = \frac{\hbar^{2}}{2 m_{\mathrm{0}} a^{2}}\cdot\frac{1}{1+y_{,1}^{2}(q^{1})} = \frac{t_{0}}{1+y_{,1}^{2}(q^{1})}
\end{align}
and the angular brackets mean an arithmetic averaging that was inserted due to the discretization of space, i.\,e.
\begin{align}
    \langle t_{\mathrm{eff}} \rangle_{i} = \frac{1}{2}\,\left[ t_{\mathrm{eff}}\left( q^{1}_{i} - \frac{a}{2} \right) + t_{\mathrm{eff}}\left( q^{1}_{i} + \frac{a}{2} \right) \right]\ .
\end{align}
With increasing the spatial resolution the model converges to the continuum theory.

\section{Transport through a Curved Wire}
\label{sec:Results}
We proceed by considering some examples of quantum scattering according to the model defined in section \ref{sec:Background}. For brevity, we set $q^{1} = x$ henceforth.


\subsection{Single Gaussian profile}
\begin{figure*}[!]
\centering
\includegraphics[width=\textwidth]{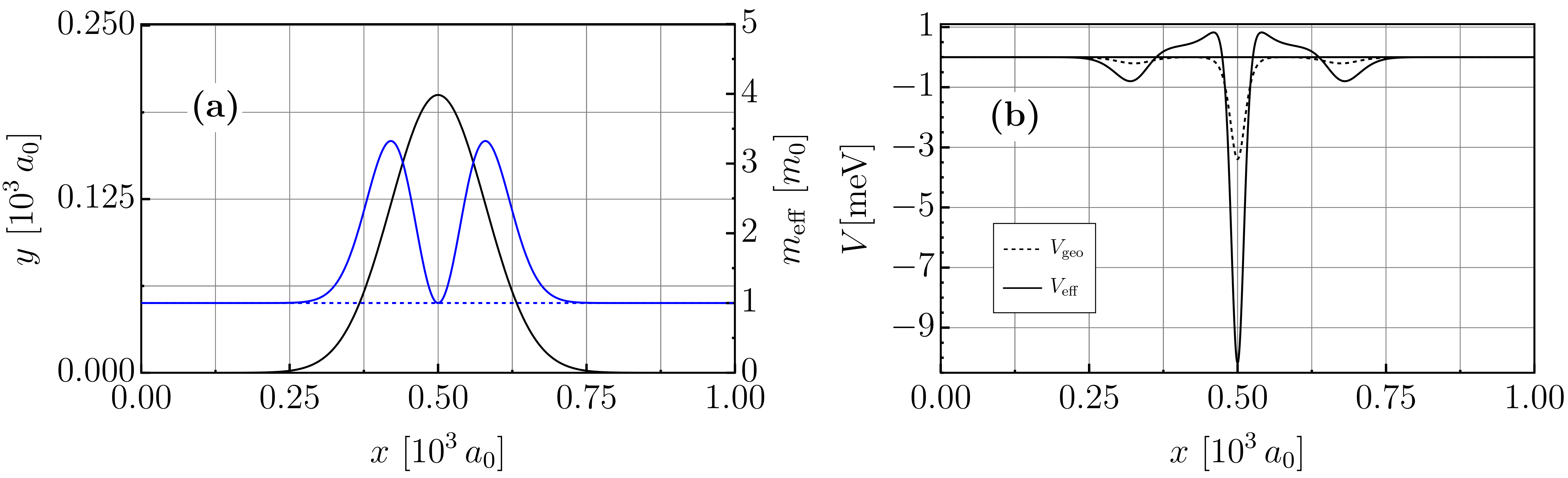}
\caption{(a) Spatial shape of the scattering region containing one Gaussian dent (solid black curve, left axis) along with the effective mass (solid blue curve, right axis). (b) Geometric and effective potentials corresponding to this situation. The parameters are $A = 0.2\,L$, $x_{0}=0.5\,L$, $\sigma = \frac{L}{4\pi}$, $m_{0} = m_{e}$, and $L = 10^{3}\,a_{0}$.}
\label{fig:Single}
\end{figure*}

\begin{figure*}[!]
\centering
\includegraphics[width=0.8\textwidth]{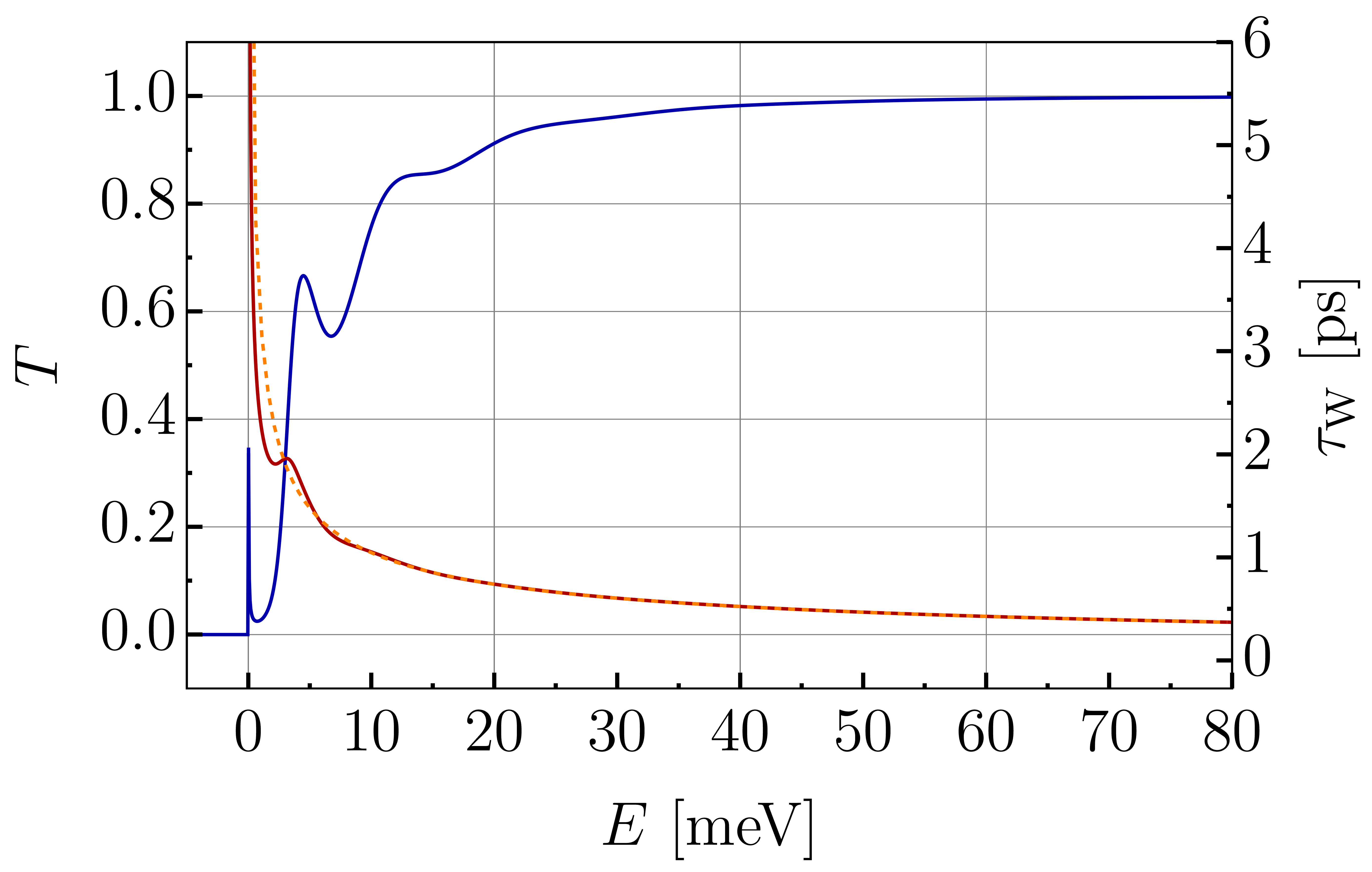}
\caption{Spectral resolved transmission coefficient (blue curve, left axis) and Wigner time delay (red curve, right axis) for a particle travelling through a quantum wire that contains a single Gaussian dent. The classical time delay due to the changed arc length is shown by the dashed orange curve. The parameters are the same as in Fig.\,\ref{fig:Single}.}
\label{fig:Scattering_Single}
\end{figure*}

We start with the quantum transport through a nanowire that is contacted by two leads in a distance of $L\in\mathbb{R}^{+}$ and shaped according to a Gaussian profile,
\begin{align}
    f(x) = A\,\ee^{-\frac{1}{2}\,\left( \frac{x-x_{0}}{\sigma} \right)^{2}}\ .
\end{align}
In our simulations, we chose $L = 10^{3}\,a_{0}$, where $a_{0}$ is the Bohr radius, and set $A = 0.2\,L$, $x_{0}=0.5\,L$, and $\sigma = \frac{L}{4\pi}$. Besides, $m_{0} = m_{e}$, the electron mass. Fig.\,\ref{fig:Single}\,(a) shows the function describing this nanowire (left axis, black curve) and how the effective mass changes with position according to \eqref{eq:effective_mass} (right axis, blue curve). Clearly, it reaches its maxima at the inflection points of the structure, corresponding to the points where the motion along the $y$-direction is the strongest pronounced. Inversely, at the maximum where the tangent is parallel to the $x$-axis, $m_{\mathrm{eff}}(x_{\mathrm{max}}) = m_{0}$ holds. The geometric potential as given by \eqref{eq:geometric_Pot} is completely attractive (this is generally the case in low dimensions \citep{Schuster2003}) and plotted in Fig.\,\ref{fig:Single}\,(b) (dashed curve). The accompanying effective potential \eqref{eq:effective_Pot} (solid curve) resulting from the transformation, however, also acts partly repulsive and possesses higher amplitudes. The overall potential landscape is evenly mirror symmetric with respect to its center.
Fig.\,\ref{fig:Scattering_Single} depicts the spectral resolved data from the transport experiment. The left axis corresponds to the blue solid curve and shows the transmission coefficient $T$ according to \eqref{eq:refelctivity+transmissivity} whereas the right one describes the red solid curve that represents the Wigner time delay \eqref{eq:Wigner_TD}. For comparison, the classical duration of motion \eqref{eq:classical_TD} is given as the dashed orange line.
\par
The transmission coefficient converges as $T \rightarrow 1\ ,\ (E \rightarrow \infty)$ and vanishes as $T \rightarrow 0\ ,\ (E \rightarrow 0)$, which is typical limiting behavior for the solutions of the one-dimensional Schrödinger equation describing a potential well \citep{Maheswari2010}. Furthermore, our transmission coefficient exhibits an oscillatory behavior hinting on pronounced reflection in the spectral range up to approximately $60\,\mathrm{meV}$. This phenomenon is attributable to the resonance states of an attractive potential \citep{Maheswari2010}, but no perfect transparency is achieved in this case of a single dent. These findings are to be compared with the familiar result for a perfectly conducting wire with the geometry of a straight line where the transmission coefficient is constantly one and the reflectivity constantly zero, correspondingly. We trace the presence of the non-ideal conducting behavior to the occurrence of scattering processes upon the geometry-induced effects, that is, the region with deviating geometric properties. Regarding Fig.\,\ref{fig:Single}\,(b) one observes that the deep negative minimum at the center of $V_{\mathrm{eff}}$ can especially support binding as an explanation thereof. Meaning, as \eqref{eq:effective_Schrodinger_effective} describes a particle travelling along the $x$-axis, the existence of at least one bound state is implied \citep{Exner1989, Goldstone1992}, resulting in the observed modification of the scattering dynamics.
\par
The Wigner time delay as it follows from the quantum computation is in line with the classical time for a wide spectral range. Below an incident particle energy of approximately $10\,\mathrm{meV}$ small deviations are visible, however. This hints on the regime where a quantum mechanical consideration is favorable and confirms at the same time that the quantum mechanically calculated Wigner time delay can indeed be interpreted as the time that the particle spends within the scattering region.

\subsection{Superposition of two Gaussian profiles}
We proceed with the consideration of two Gaussian dents placed behind each other. That is, the graph is described by the functions
\begin{align}
    f_{s}^{\pm}(x) = A\,\left( \ee^{-\frac{1}{2}\,\left( \frac{x-x_{0} + s}{\sigma} \right)^{2}} \pm \ee^{-\frac{1}{2}\,\left( \frac{x- x_{0} - s}{\sigma} \right)^{2}} \right)\ ,
\end{align}
where $s\in\mathbb{R}$ measures the shift of the single dents from the position $x_{0}\in\mathbb{R}^{+}$ of the scattering interval, and the remaining parameters are the same as in the previous section. The resulting structures possess a reflection center there, and the $\pm$ sign indicates whether they are asserted even $(+)$ or odd $(-)$ parity with respect to this transformation, meaning if the $y$-displacement has the same or the opposite course for both dents, respectively.
\par
\begin{figure*}[!]
\centering
\includegraphics[width=0.9\textwidth]{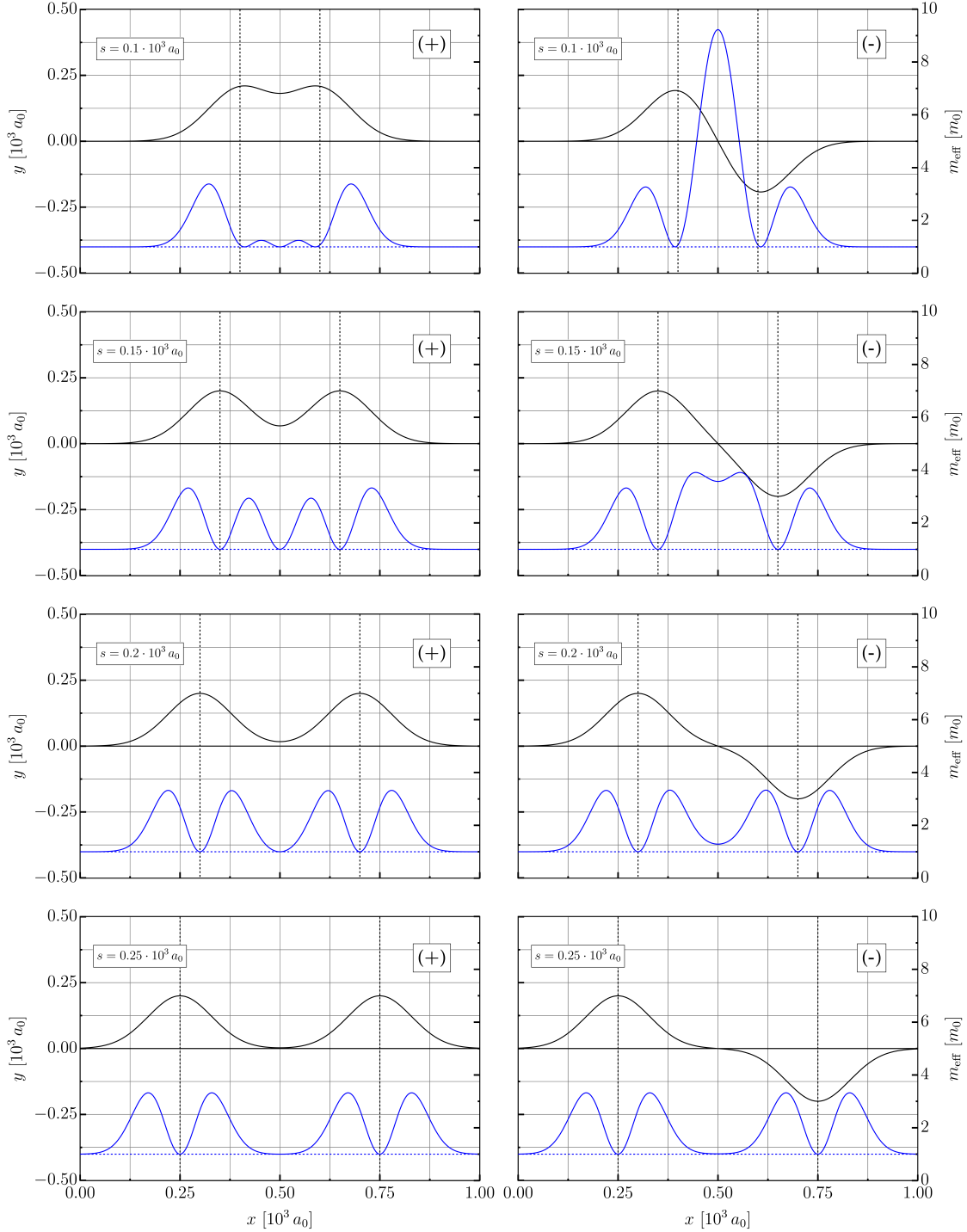}
\caption{Spatial shapes of the scattering regions containing two Gaussian dents (solid black curve, left axes) along with the effective masses (solid blue curves, right axes). The parameters are the same as in Fig.\,\ref{fig:Single}, with the value of $s$ being indicated on the figures. The relative distance of the dents is $2s$. The left and right columns correspond to the configurations with even and odd parities, respectively.}
\label{fig:Setting_Double}
\end{figure*}
\begin{figure*}[!]
\centering
\includegraphics[width=\textwidth]{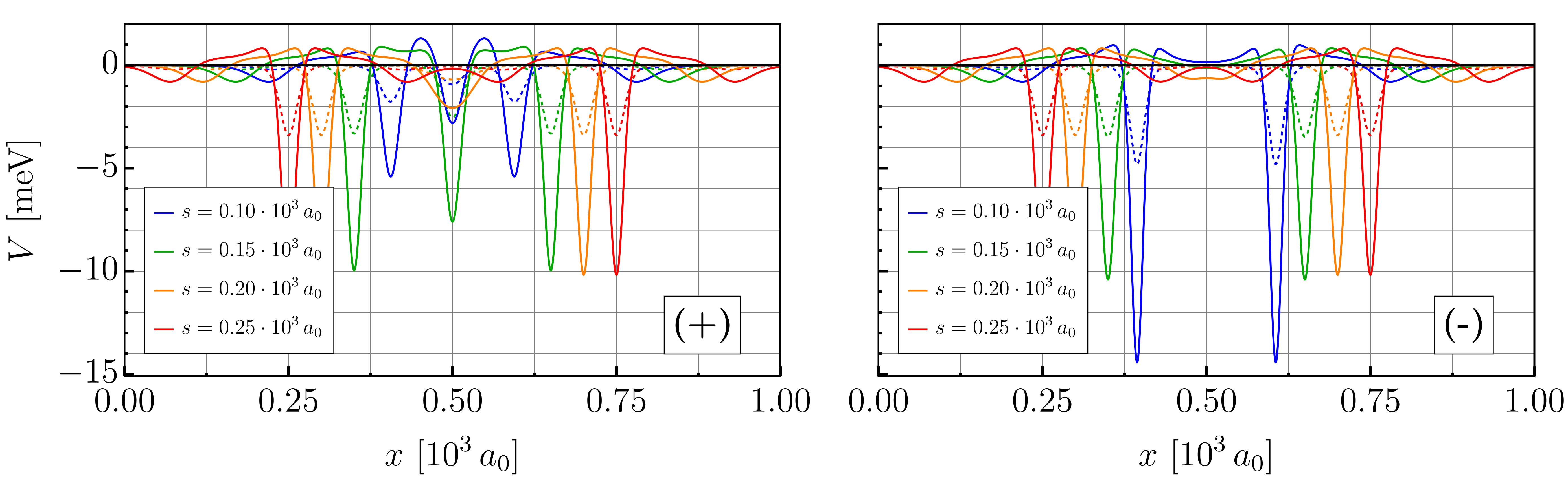}
\caption{Geometric (dashed curves) and effective (solid curves) potential fields for the parity even (left) and odd (right) double dent structures. The parameters are the same as in Fig.\,\ref{fig:Single}, with only $s$ varying.}
\label{fig:Potentials_Double}
\end{figure*}
\begin{figure*}[!]
\centering
\includegraphics[width=0.9\textwidth]{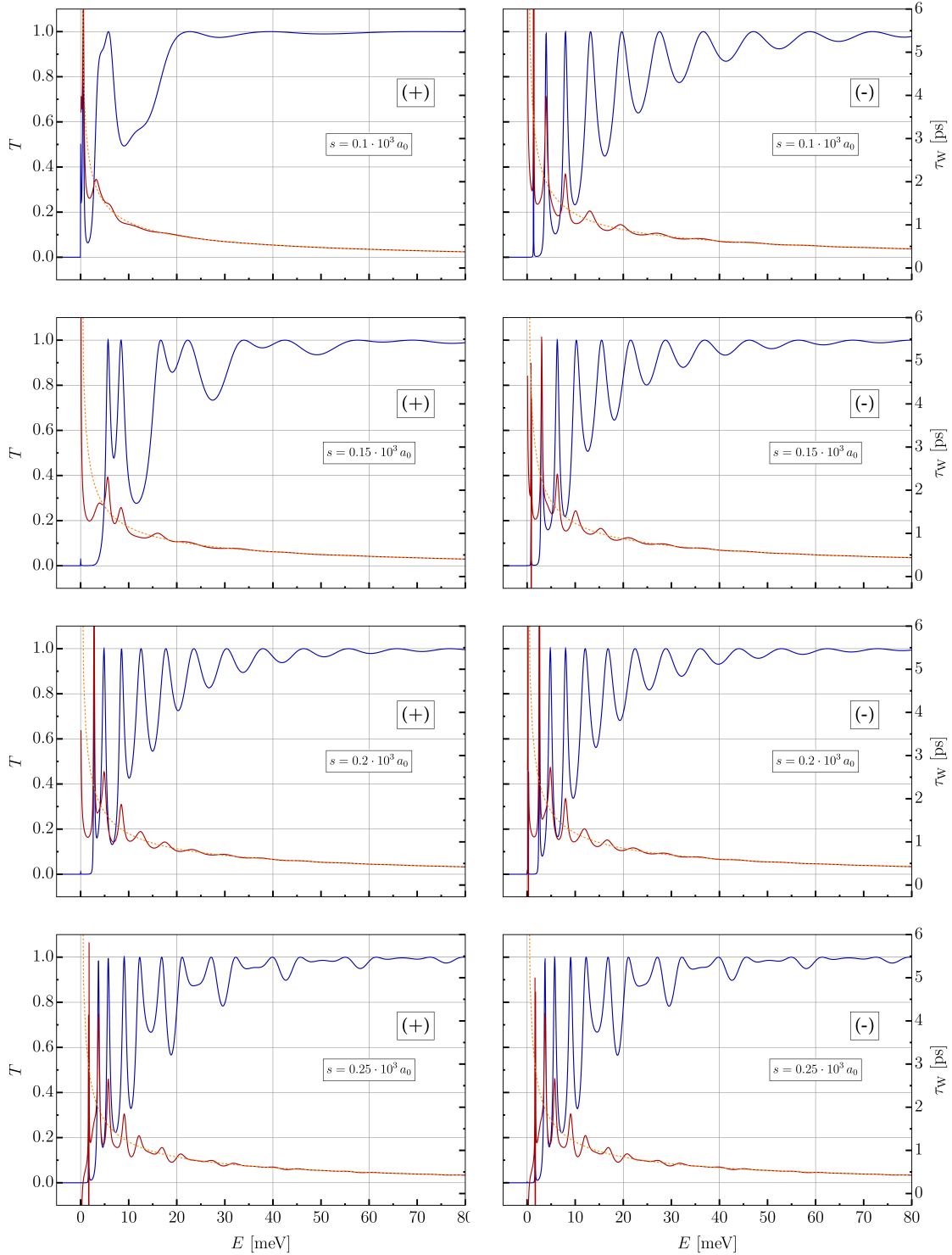}
\caption{Spectral resolved transmission coefficient (blue curves, left axes) and Wigner time delay (red curves, right axes) for a quantum particle travelling along a one-dimensional manifold that contains two Gaussian dents. The classical time delays due to the changed arc length are shown by the dashed orange curves. The left and right columns correspond to the configurations with even and odd parities, respectively.}
\label{fig:Scattering_Double}
\end{figure*}

Fig.\,\ref{fig:Setting_Double} shows the profiles of the wire and the corresponding effective masses for both cases as the parameter $s$ varies. The position of the peaks of the single dents is marked  by dashed lines. Fig.\,\ref{fig:Potentials_Double} complements the potential fields. The plots show that both the effective masses and the effective potentials are almost identical when the two dents in the structure are sufficiently far separated from each other, i.\,e., when their overlap is negligible like for $s \geq 0.25\,L$. This is because the geometric properties upon which the Hamilton operator $\hat{T}[g,\kappa]$ in \eqref{eq:effective_Schrodinger_effective} depends are not sensitive to the orientation of the structure within the ambient space. Consequently, as given in Fig.\,\ref{fig:Scattering_Double}, the scattering data is also indistinguishable for two respectively shaped wires. However, the smaller $s$ becomes the stronger the dents overlap, and both the effective mass and the effective potential adapt. Eventually, we observe significant differences between the two cases with opposite parities (compare figs.\,\ref{fig:Setting_Double} and \ref{fig:Potentials_Double} for $s \leq 0.2\,L$).
\par
Fig.\,\ref{fig:Scattering_Double} presents the corresponding results for the transmission coefficient $T$ and Wigner time delay $\tau_{\mathrm{W}}$. A common feature of both configurations in the course of $T$ is the appearance of resonances indicating perfect transparency of the scattering region for a discrete set of energies within the observed spectral range. This is to be contrasted with the results for a single dent (see Fig.\,\ref{fig:Scattering_Single}). Following  the well-established one-dimensional short range scattering theory \citep{Maheswari2010}, we explain the occurrence of the resonances in $T$ by the formation of standing waves within the scattering region. The mean lifetimes of the resonances may be defined by the mean peak widths. Again, we observe the vanishing of the transmission coefficient, $T \rightarrow 0$\ for $E \rightarrow 0$, indicating the existence of bound states \citep{Maheswari2010, Exner1989, Goldstone1992}.
\par
The Hamilton operator \eqref{eq:effective_Schrodinger_effective} depends on the geometric properties of the wire in a nonlinear way. Therefore, scattering quantities  are sensitive to small structural changes. The aforementioned parity-related difference in the effective masses and potentials results in considerably different qualitative features in the spectral dependencies of the transmission coefficient, even when the individual dents are still well distinguishable. For example, the situation for $s = 0.15\,L$ shows that the even configuration tends to possess less resonances than the odd one, and that their spectral distribution is different. The Wigner time delays, however, do not differ crucially for both cases and align well with the estimation based on classical arguments, except for low energies where their courses are disturbed. These deviations mean an effect of the quantum scattering processes on the time that it takes to measure the particle and thus underpin their existence.
\par
A further remark concerns the Hartman effect which, in conventional potential scattering of waves, refers to a saturation of the delay (or tunneling) time with the width of the potential. In our case for scattering centers given by the metric tensor (which are predominant in the high energy region, compare \citep{Meschede2023}), the Hartman effect is absent. This directly follows from the fact that our classical and quantum delay times agree well, implying that there is no tunneling, only a time lag depending on the arc length. This behavior is a direct consequence of the particularities of a quantum particle moving freely along a non-Euclidean space. In the low energy region or in the case when the particle is subject to a scalar potential, we expect a corresponding contribution to the time delay and there is no reason to exclude the occurrence of a Hartman effect from this one.
%

\section{Conclusions}
\label{sec:Conclusion}
We studied the motion of a quantum particle moving freely along one-dimensional Riemannian manifolds with nontrivial geometry, embedded in an otherwise structureless hyperspace. We showed that even for a free particle the Hamilton operator is a functional of the geometric invariants of the underlying space, leading to quantum scattering. Numerically, we performed and presented transport experiments through planar curves whose profiles contain localized Gaussian dents. For these, we evaluated the transmission coefficient and the Wigner time delay. The results differ significantly from the familiar ones found for an ideal wire with the shape of a straight line. These features can be attributed to mainly attractive effective potentials that we derived explicitly as functions of the geometric parameters of the manifold, as we assert them to the emergence of both bound and resonance states. The Wigner delay time was put into perspective with the classical motion and conventional scalar potential scattering.

\section{Acknowledgement}
This work is supported by the Deutsche Forschungsgemeinschaft (DFG) under  project number 429194455.

\FloatBarrier
\bibliography{scatteringTD}

\end{document}